\documentclass{article}

\usepackage{arxiv}

\usepackage[utf8]{inputenc} 
\usepackage[T1]{fontenc}    
\usepackage{hyperref}       
\usepackage{url}            
\usepackage{booktabs}       
\usepackage{amsfonts}       
\usepackage{nicefrac}       
\usepackage{microtype}      
\usepackage{lipsum}		
\usepackage{graphicx}

\usepackage{subfigure}
\usepackage{multirow}
\usepackage{mathtools,amssymb}
\usepackage{stmaryrd} 

\title{PT-Ranking: A Benchmarking Platform for Neural Learning-to-Rank}

\author{ \href{https://orcid.org/0000-0002-1569-8507}{\includegraphics[scale=0.06]{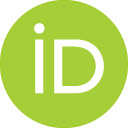}\hspace{1mm}Hai-Tao Yu}\\
	Faculty of Library, Information and Media Science\\
	University of Tsukuba\\
	1-2 Kasuga, Tsukuba City, Ibaraki, 305-8550, Japan\\
	\texttt{yuhaitao@slis.tsukuba.ac.jp}
}

\renewcommand{\shorttitle}{\textit{arXiv} Template}

\hypersetup{
pdftitle={A template for the arxiv style},
pdfsubject={q-bio.NC, q-bio.QM},
pdfauthor={David S.~Hippocampus, Elias D.~Striatum},
pdfkeywords={First keyword, Second keyword, More},
}

\begin{document}
\maketitle

\begin{abstract}
Deep neural networks has become the first choice for researchers working on algorithmic aspects of learning-to-rank. Unfortunately, it is not trivial to find the optimal setting of hyper-parameters that achieves the best ranking performance. As a result, it becomes more and more difficult to develop a new model and conduct a fair comparison with prior methods, especially for newcomers. In this work, we propose \textit{PT-Ranking}\footnote{https://github.com/wildltr/ptranking}, an open-source project based on PyTorch for developing and evaluating learning-to-rank methods using deep neural networks as the basis to construct a scoring function. On one hand, PT-Ranking includes many representative learning-to-rank methods. Besides the traditional optimization framework via empirical risk minimization, adversarial optimization framework is also integrated. Furthermore, PT-Ranking's modular design provides a set of building blocks that users can leverage to develop new ranking models. On the other hand, PT-Ranking supports to compare different learning-to-rank methods based on the widely used datasets (e.g., MSLR-WEB30K, Yahoo!LETOR and Istella LETOR) in terms of different metrics, such as precision, MAP, nDCG, nERR. By randomly masking the ground-truth labels with a specified ratio, PT-Ranking allows to examine to what extent the ratio of unlabelled query-document pairs affects
the performance of different learning-to-rank methods. We further conducted a series of demo experiments to clearly show the effect of different factors on neural learning-to-rank methods, such as the activation function, the number of layers and the optimization strategy. The experimental results reveal that the aforementioned factors significantly affect the final performance. Careful examinations of these factors are highly recommended. To summarize, PT-Ranking is highly complementary to the previous open-source projects for learning-to-rank. We envision that PT-Ranking will lower the technical barrier and provide a convenient open-source platform for evaluating and developing learning-to-rank models in different fields, and thus facilitate researchers from various backgrounds.
\end{abstract}


\section{Introduction}
\label{sec:Intro}
Learning-to-rank has been intensively studied and has shown significantly increasing values in a wide range of domains, such as \textit{web search}, \textit{recommender systems}, \textit{dialogue systems}, \textit{machine translation}, \textit{computer vision} and even \textit{computational biology}, to name a few. The information retrieval (IR) community has experienced a flourishing
development of learning-to-rank methods, such as \textit{pointwise} methods, \textit{pairwise} methods and \textit{listwise} methods. The pointwise
methods \cite{SubsetRegression, RameshL2R, ChuGaussian, ChuSVOR} transform the ranking problem into
a task of (ordinal) regression or classification on individual documents. A major problem
is that the pointwise methods are agnostic to the relevance-based
order information among documents that are associated with the same
query. To make a step forward, the pairwise methods \cite{FreundBoosting, ShenPerceptron, RankSVMStruct}
were then proposed, which transform the ranking problem into a task
of pairwise classification. However, the loss functions merely consider
the relative order between two documents rather than the total order
relationship among all documents associated with the same query. Moreover, the number of document pairs per
query may differ from query to query, thus the result can be biased
in favor of queries with more documents in the training data \cite{RankCosine}. To overcome the shortcomings of the aforementioned two categories
of ranking methods, the listwise methods \cite{OlivierMargin, AdaRank, YueSVMAP, GuiverSoftRank, SoftRank, QinApproximateNDCG, LambdaMART, ListNet, ListMLE, BoltzRank, LambdaRank, RankCosine} appeal to the loss function that is defined over all documents associated with the same query. Recently, inspired by generative adversarial network (GAN) and its variants, significant efforts \cite{IRGAN, AdverPRR, AdverSMIR, AdverCMR, AdverPreference, AdverFGIMR} have been made to develop meaningful adversarial optimization methods for addressing learning-to-rank problems.

Despite the success achieved by the aforementioned methods for learning-to-rank, there are still many open issues. On one hand, with the recent advances in machine learning, the learning-to-rank models are getting increasingly complex. Take the ranking methods based on neural networks for example, it is not trivial to find the optimal setting of hyper-parameters that achieves the best ranking performance. As a result, it becomes more and more difficult to develop a new model and conduct a fair comparison with prior methods, especially for newcomers. On the other hand, the recent publications \cite{WorryingAnalysis, MetricRealityCheck, NeuralHype} pointed out that some reported improvements \textit{don't add up}. The factors that contribute to such phenomena include: (1) using weak baseline methods; (2) difficulties in comparing or reproducing results across papers; (3) using various types of datasets, performance measures and data preprocessing steps. Hence, both academia and industry have recognized the critical importance and the long-term value in developing and maintaining open source projects on popular research topics, such as learning-to-rank. This is also why the \textit{replicability} and \textit{reproducibility} of published experiments have gained increasing attention in the IR research community, as evidenced by the recent workshop \cite{2015RIGOR}, the annual reproducibility track since 2015 of the European Conference on Information Retrieval (ECIR)  and the Association for Computing Machinery (ACM) policy on \textit{Artifact Review and Badging}\footnote{https://www.acm.org/publications/policies/artifact-review-badging} in computer science. 

Motivated by the aforementioned open issues, the focus of this paper is on developing a benchmarking platform for learning-to-rank methods based on neural networks, which is referred to as PT-Ranking. PT-Ranking is implemented as a lightweight Python library based on PyTorch. It can be used within a JupyterLab notebook, where users can make use of the inline plots and interactive visualization features. The main contributions are summarized as follows:\\
\indent \textbf{(1)} PT-Ranking includes a large number of representative learning-to-rank methods, such as the pointwise method RankMSE (ranking based on least mean squares regression \cite{PRML}) , pairwise methods \cite{RankingSVM, MSnDCGRankNet} and listwise methods \cite{TaoWSDM2019, ListMLE, LambdaRank, RankCosine, ListNet, QinApproximateNDCG, StochasticTreatmentRF}. Moreover, besides the traditional optimization strategy via empirical risk minimization, PT-Ranking also includes pointwise, pairwise and listwise methods based on adversarial optimization  \cite{IRGAN}, which enables to pinpoint the pros and cons of different optimization frameworks. In order to make a fair comparison with the state-of-the-art approach LambdaMART that builds upon the technique of gradient boosting decision tree (GBDT), the implementations of LambdaMART provided in LightGBM \cite{LightGBM} and XGBoost \cite{XGBoost} are also included.\\
\indent \textbf{(2)} PT-Ranking supports to compare different learning-to-rank methods based on the widely used datasets (e.g., MSLR-WEB30K, Yahoo!LETOR and Istella LETOR) in terms of different metrics, such as precision, MAP, nDCG, nERR. By randomly masking the ground-truth labels with a specified ratio, PT-Ranking allows to examine to what extent the ratio of unlabelled query-document pairs affects
the performance of different learning-to-rank methods.\\
\indent \textbf{(3)} PT-Ranking offers deep neural networks as the basis to construct a scoring
function. On one hand, PT-Ranking provides facilities to investigate the effects of different hyper-parameters, such as activation functions and number of layers. On the other hand, the simplified modules make it very easy to examine a new loss function or a new optimization strategy. Thanks to this, PT-Ranking facilitates the understanding, comparing, designing of learning-to-rank methods.

The remainder of the paper is structured as follows. In the next section, we briefly survey the existing open-source projects which are related to learning-to-rank. In section 3, we give the mathematical formulation of two different learning-to-rank frameworks following the Cranfield paradigm. In section 4, we detail the key components of PT-Ranking for learning-to-rank. In section 5, we demonstrate PT-Ranking's functionalities through a series of demo experiments based on benchmark datasets. Finally, we conclude the paper in section 6.

\section{Related Work}
\label{sec:ReWork}
In this section, we discuss the existing open-source projects on learning-to-rank and show what primary facilities they offer.

RankLib\footnote{http://www.lemurproject.org/ranklib.php} is a Java package that implements eight popular learning-to-rank methods, as well as several evaluation metrics. Unfortunately, due to the platform limitation, it is not easy to customize and/or further extend some pre-implemented models, especially in using deep neural networks as the basis to construct a scoring function. 

QuickRank\footnote{http://quickrank.isti.cnr.it}, RankEval \cite{RankEval}, XGBoost \cite{XGBoost}, LightGBM \cite{LightGBM}, CatBoost \cite{CatBoost} are the leading packages focusing on tree-based models. A representative implementation is the LambdaMART method \cite{LambdaMART} which builds upon gradient boosted decision trees. QuickRank introduces post-learning optimisations pipelined with the learning-to-rank methods. RankEval allows to conduct a structural analysis reporting statistics about shape, depth and balancing of trees in the forest. However, the tree-based models commonly require extensive feature engineering in order to handle textual features. Moreover, we note that XGBoost, LightGBM and CatBoost do not provide dedicated functionalities for learning-to-rank, in terms of both algorithms and metrics. 

Due to the breakthrough successes of neural networks, many approaches \cite{DSSM, CDSSM, DRMM, HuCNNMatching, PangAAAMatching, MatchSRNN} building upon neural networks are proposed, which are referred to as neural ranking models. Different from the tree-based models that require extensive feature engineering to handle textual features. Neural ranking models can effectively handle sparse features through embeddings. Recently, a number of open-source projects, such as TF-Ranking \cite{TFRanking} and MatchZoo \cite{MatchZoo}, have emerged, which builds upon either TensorFlow or PyTorch. We note that MatchZoo focuses on text matching research. The typical tasks are question answer, information retrieval, and textual entailment. The benchmark datasets, such as MSLR-WEB30K and Yahoo!LETOR, are not supported. Though TF-Ranking supports LETOR datasets in LibSVM format, a number of representative learning-to-rank methods are not included, especially the methods based on adversarial optimization \cite{IRGAN}. Another possible barrier is that some researchers prefer to use PyTorch rather than TensorFlow. 

PT-Ranking is highly complementary to the aforementioned open-source projects. Yet, to the best of our knowledge, we are the first to support an in-depth comparison of many representative learning-to-rank methods based on PyTorch across several benchmark datasets, such as MSLR-WEB30K and Yahoo!LETOR.
\section{Learning-to-Rank}
\label{sec:L2R}
In this section, we describe the general learning-to-rank formulation following the Cranfield paradigm, where two different optimization frameworks are introduced.
\subsection{Preliminaries}
\label{subsec:pre}
Let $\mathcal{Q}$ and $\mathcal{D}$ be the query space and the document
space, respectively. We use $\Phi:\mathcal{Q}\times\mathcal{D}\rightarrow\mathcal{Z}\coloneqq\mathbb{R}^{d}$
to denote the mapping function for generating a feature vector for
a document under a specific query context, where $\mathcal{Z}$ represents
the $d$-dimensional feature space. We use $\mathcal{T}\coloneqq\mathbb{R}$
to denote the space of the ground-truth labels each document
receives. Thus for each query, we have a list of document feature
vectors $\mathbf{x}=(x_{1},...,x_{m})\in\mathcal{X}\coloneqq\mathcal{Z}^{m}$
and a corresponding list $\mathbf{y}^{\ast}=(y_{1}^{\ast},...,y_{m}^{\ast})\in\mathcal{Y}\coloneqq\mathcal{T}^{m}$
of ground-truth labels. The subscript $i$ like $x_{i}$ or $y_{i}^{\ast}$
denotes the $i$-position in the list. In practice, we get independently
and identically distributed (i.i.d) samples $\mathcal{S}=\{(\mathbf{x}_{j},\mathbf{y}_{j}^{\ast})\}_{j=1}^{n}$
from an unknown joint distribution $P(\cdot,\cdot)$ over $\mathcal{X}\times\mathcal{Y}$. A ranking $\pi$ on $m$ documents $\mathbf{x}=(x_{1},...,x_{m})$ is defined as a permutation of $\mathbf{x}$. $\pi(i)$ / $\pi(x_i)$ yields the \textit{rank} of the $i$-th document within $\mathbf{x}$. $\pi^{-1}(r)$ yields the
index within $\mathbf{x}$ of the document at rank $r$, and we have $\pi^{-1}(\pi(i))=i$ or $\pi^{-1}(\pi(x_i))=i$. Since we are interested in sorting documents in descending order according to their relevance, we think of higher positions with smaller rank values as more favorable. A ground-truth ranking refers to the ideal ranking of documents that are sorted according to their real relevance to the query under consideration. We note that there are multiple ideal rankings for a query when we use graded relevance labels due to label ties. We use $f:\mathbf{x}\rightarrow\mathbb{R}^{m}$ to denote the real-valued scoring function,
which assigns each document a score. One can design various ranking methods by deploying different loss functions to learn the parameters $\theta$ based on the training data. In the testing phase, the scores of the documents associated with the same query, i.e., $\mathbf{y}=f(\mathbf{x})=(f(x_{1}),f(x_{2}),..., f(x_{m}))$, are used to sort the documents.
\subsection{Empirical Risk Minimization}
\label{subsec:ERM}
Typically, we measure the loss of ranking documents for a query using $f$ with
a loss function $\mathcal{R}(f(\mathbf{x}),\mathbf{y}^{\ast})$, which is commonly rank-sensitive. Then the goal of learning-to-rank is to learn the optimal scoring function over a hypothesis space $\mathcal{F}$ of ranking functions that can \emph{minimize the expected risk} as defined below:
\begin{equation}
\min_{f\in\mathcal{F}}\Re(f)=\min_{f\in\mathcal{F}}\int_{\mathcal{X}\times\mathcal{Y}}\mathcal{R}(f(\mathbf{x}),\mathbf{y}^{\ast})dP(\mathbf{x},\mathbf{y}^{\ast})
\end{equation}

Because $\Re(f)$ is intractable to optimize directly and the joint distribution is unknown, we appeal to the \emph{empirical risk minimization} to approximate the expected risk, which is defined as follows:
\begin{equation}
\label{eq:erm}
\min_{f\in\mathcal{F}}\tilde{\Re}(f;\mathcal{S})=\min_{f\in\mathcal{F}}\frac{1}{n}\sum_{j=1}^{n}\mathcal{R}(f(\mathbf{x}_j),\mathbf{y}^{\ast}_j)
\end{equation}

Most learning-to-rank methods of this kind  differ primarily in how they define the surrogate loss function $\mathcal{R}$. These methods are grouped into three categories: pointwise methods \cite{SubsetRegression, RameshL2R, ChuGaussian, ChuSVOR}, pairwise methods \cite{FreundBoosting, ShenPerceptron, RankSVMStruct} and listwise methods \cite{OlivierMargin, AdaRank, YueSVMAP, GuiverSoftRank, SoftRank, QinApproximateNDCG, LambdaMART, ListNet, ListMLE, BoltzRank, LambdaRank, RankCosine}.

\subsection{Adversarial Optimization}
\label{subsec:AdOpt}
Inspired by \cite{GAN, IRGAN}, we can formulate the process of learning-to-rank as a game between two opponents: a \textit{generator} and a \textit{discriminator}. The generator aims to generate (or select) rankings
that look like the ground-truth ranking, which may fool the discriminator. Whereas the discriminator aims to
make a clear distinction between the ground-truth ranking and the
ones generated by its opponent generator. The framework for adversarial learning-to-rank is given as:
\begin{equation}
\label{Eq:AL2R}
J^{G^{*},D^{*}}=\min_{\theta}\max_{\phi}\sum_{n=1}^{N}\mathbb{E}_{\pi\backsim P_{true}(\pi|q_{n})}[\log D_{\phi}(\pi|q_{n})]+\mathbb{E}_{\pi\backsim P_{\theta}(\pi|q_{n})}[\log(1-D_{\phi}(\pi|q_{n}))]
\end{equation}
where the generator $G$ is denoted as $P_{\theta}(\pi|q_{n})$ that aims to minimize
the objective. On one hand, the generator fits the true distribution
over all possible rankings $\pi\backsim P_{true}(\pi|q)$. On the
other hand, it randomly generates rankings in order to fool the discriminator. 
The discriminator is denoted as $D_{\phi}(\pi|q_{n})$,
which estimates the probability of a ranking being either the ground-truth
ranking or not. The objective of the discriminator is to maximize the log-likelihood
of correctly distinguishing the ground-truth ranking from artificially
generated rankings. Furthermore, we are able to perform adversarial learning in a pointwise ($k=1$), pairwise ($k=2$) and listwise manner ($k\gg1$), respectively by adjusting the size of ranking. 

For adversarial learning-to-rank, both generator and discriminator are designed to be scoring functions. In particular, instead of generating new document feature vectors, the generation of rankings via generator is formulated as a sampling process. Due to space limitation,  we refer readers to the paper \cite{IRGAN} for the details on how to optimise generator and discriminator.
\section{Platform Overview}
\label{sec:ListwiseMiniMax}

In the following, we first show how to develop a new learning-to-rank model based on PT-Ranking. Then we detail its key components.

\begin{figure}[!htbp]	
	\centering	
	\includegraphics[width=5.3in, totalheight=3.1in]{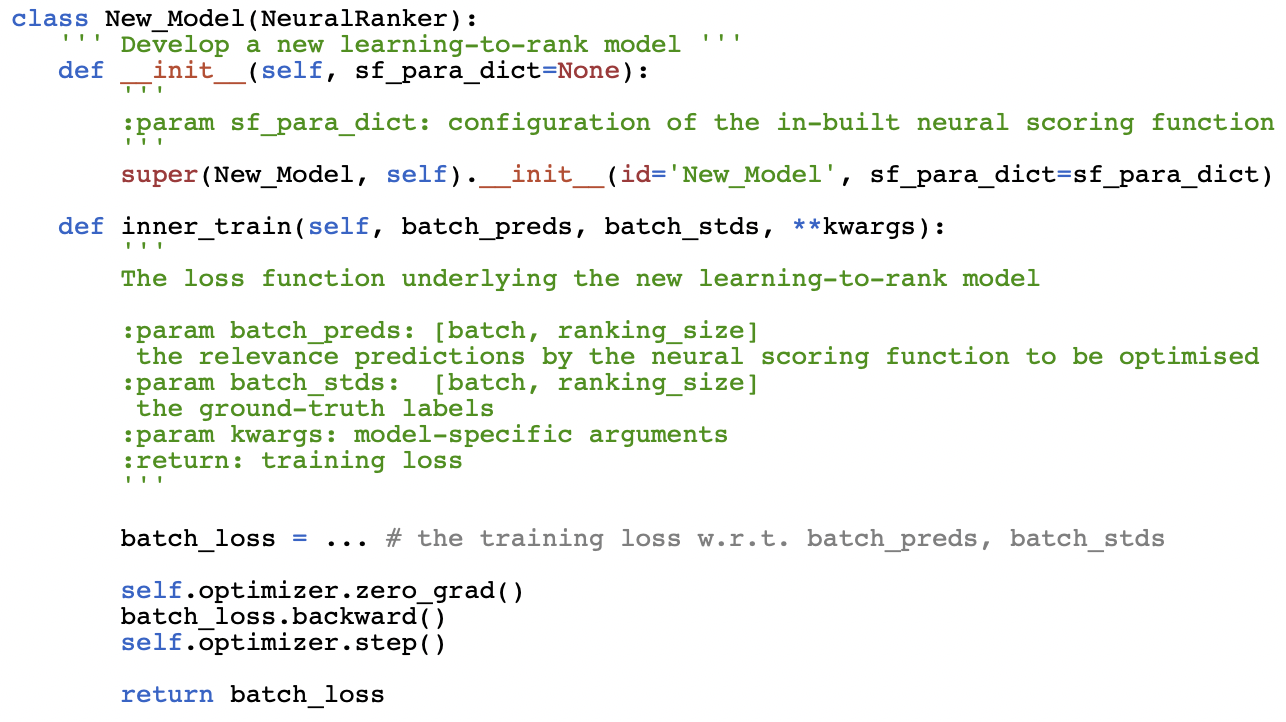}
	\caption{Develop a new learning-to-rank model.}
	\label{Fig:newltrmodel}
\end{figure}

PT-Ranking offers deep neural networks as the basis to construct a scoring function based on PyTorch and can thus fully leverage the advantages of PyTorch. NeuralRanker is a class that represents a general learning-to-rank model. A key component of NeuralRanker is the neural scoring function $f$. The configurable hyper-parameters include activation function, number of layers, number of neurons per layer, etc. All specific learning-to-rank models inherit NeuralRanker and mainly differ in the way of computing the training loss $\mathcal{R}$. Figure \ref{Fig:newltrmodel} shows the main step in developing a new learning-to-rank model following Eq-\ref{eq:erm}, where batch\_preds and batch\_stds correspond to $f(\mathbf{x})$ and $\mathbf{y}^{\ast}$, respectively. We can observe that the main work is to define the surrogate loss function $\mathcal{R}$.

\begin{figure}[!htbp]
	\centering
	\includegraphics[width=5.8in, totalheight=1.5in]{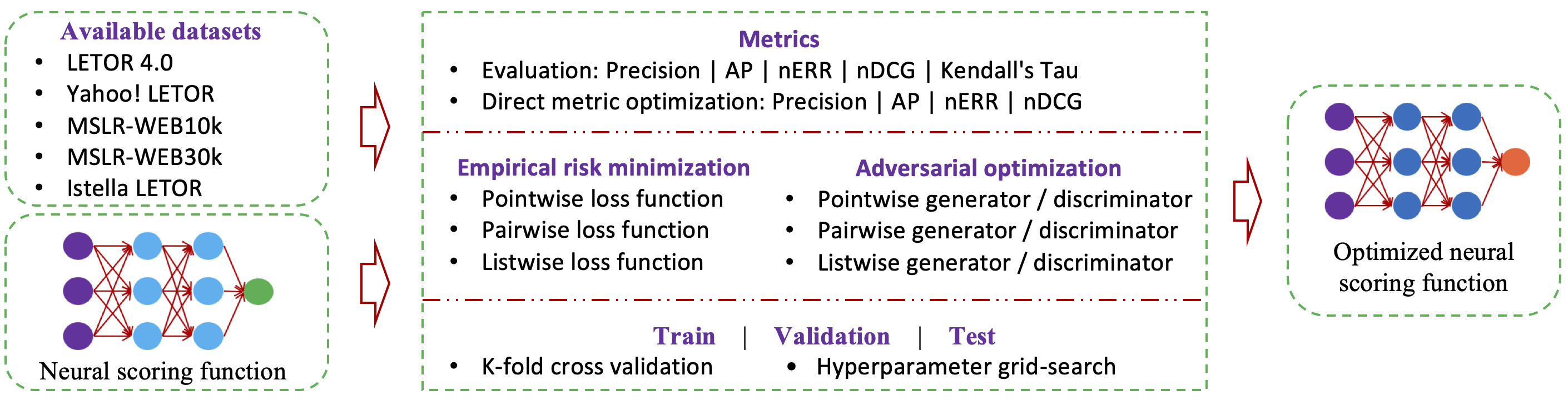}	
	\caption{The overall architecture of PT-Ranking.}
	\label{fig:ptr}	
\end{figure}

Figure \ref{fig:ptr} illustrates the overall architecture of PT-Ranking. The currently supported datasets are LETOR4.0 \cite{LETORIR}, Yahoo! LETOR \cite{YahooL2RData}, MSLR-WEB10K, MSLR-WEB30K\footnote{https://www.microsoft.com/en-us/research/project/mslr/} and Istella LETOR\footnote{http://quickrank.isti.cnr.it/istella-dataset/}. For more detailed information, e.g., the feature description, we refer readers to the corresponding papers. When loading a specified dataset, the supported functionalities are: (1) Label binarization, namely binarize the ground-truth labels if needed; (2) Random masking with a specified ratio, i.e., randomly mask the ground-truth labels per query as unlabelled ones; (3) Feature normalization. For datasets that are provided with raw features, such as MSLR-WEB10K AND MSLR-WEB30K, different methods for query-level normalization are provided. PT-Ranking supports the widely used evaluation metrics, such as Precision, Average Precision (AP), Normalized Discounted Cumulative Gain (nDCG) \cite{FstnDCG} and Expected Reciprocal Rank (ERR) \cite{ERR}. On one hand, these metric can be used to measure the performance of learning-to-rank methods. On the other hand, PT-Ranking also includes methods on how to directly optimize these metrics.

Given the configured neural scoring function, we can choose different models and different optimization frameworks (detailed in Section \ref{sec:L2R}) to learn its parameters. We can also examine the effects of different hyper-parameters, namely, grid-search over hyper-parameters of a specific model.

\section{DEMO Experiments}
In this section, we show PT-Ranking's functionalities through a series of demo experiments, namely in-depth comparisons of different learning-to-rank models based on benchmark datasets. In our experiments, we used the publicly available datasets, MSLR-WEB10K and MSLR-WEB30K, where each query-document pair is represented with a feature vector. The ground truth is a multiple-level relevance judgment, which takes $5$ values from $0$ (irrelevant) to $4$ (perfectly relevant). We use nDCG to measure the performance. We report the results with different cutoff values $1$, $3$, $5$, $10$, $20$ and $50$ to show the performance of each method at different positions.
\subsection{Methods}
For traditional learning-to-rank via empirical risk minimization, a number of typical methods are adopted. RankMSE is a simple pointwise method. RankNet \cite{MSnDCGRankNet} represents the pairwise method. The listwise methods include ListNet \cite{ListNet}, ListMLE \cite{ListMLE}, RankCosine \cite{RankCosine},  LambdaRank \cite{LambdaRank}, ApproxNDCG \cite{QinApproximateNDCG}, WassRank \cite{TaoWSDM2019} and ST-ListNet \cite{StochasticTreatmentRF}. Specifically, for ListNet, the ranking loss is computed based on the top-1 approximation as in the original paper \cite{ListNet}, namely each element of the probability vector represents the probability of the corresponding document being ranked at the top-1 position. For WassRank, the suggested parameter configuration by \cite{TaoWSDM2019} is used. Following the recent studies \cite{RevisitingApproxNDCG, StochasticTreatmentRF}, for ApproxNDCG, the parameter $\alpha$ is set as $10$. Given the raw features per query-document pair, they are normalized using the \textit{z-score} method at a query level. We further use batch normalization between consecutive layers. 

For adversarial learning-to-rank, IRGAN \cite{IRGAN} is implemented to represent the main approach that adversarially optimizes scoring functions for ranking. The pointwise and pairwise versions are denoted as IRGAN-Point and IRGAN-Pair, respectively. The temperature is set as $0.5$. We note that how to perform adversarial learning-to-rank in a listwise manner is not resolved in \cite{IRGAN}. To address this issue, we formulate both generator and discriminator with the Plackett-Luce model \cite{PlackettLuceModel-2}, namely
\begin{equation}
P_{\theta}(\pi|q_{n})=\prod_{i=1}^{m}\frac{\exp(f_{\theta}(x_{\pi^{-1}(i)}))}{\sum_{j=i}^{m}\exp(f_{\theta}(x_{\pi^{-1}(j)}))}
\end{equation}
\begin{equation}
D_{\phi}(\pi|q_{n})=\prod_{i=1}^{m}\frac{\exp(f_{\phi}(x_{\pi^{-1}(i)}))}{\sum_{j=i}^{m}\exp(f_{\phi}(x_{\pi^{-1}(j)}))}
\end{equation}

Inspired by the work of Bruch et al. \cite{StochasticTreatmentRF}, we resort to the Gumbel-softmax trick \cite{GumbelSoftmax, ConcreteDistribution}  in order to enhance the efficiency of sampling rankings with $f_{\theta}$. Specifically, we associate an i.i.d sample drawn from $Gumbel(0,1)$ to each document for the query under consideration (i.e., $\mathbf{g}=g_{1},...,g_{m}$ for $\mathbf{x}=x_{1},...,x_{m}$). We then sort $\hat{\mathbf{y}}=\mathbf{g}+f_{\theta}(\mathbf{x})$ in an decreasing order. The corresponding re-ranking of $\mathbf{x}$ is regarded as a sample ranking of the generator. We refer to the method as IRGAN-List. We also test two different values of ranking size, $5$ and $10$, and the corresponding methods are denoted as IRGAN-List-5 and IRGAN-List-10, respectively. For all the methods, the inner loop for training both generator and discriminator is set as $1:1$. We used a simple $5$-layer feed-forward neural network to approximate the scoring function, where the size of a hidden layer is set as $100$. According to the studies \cite{IRGAN, AdverSMIR}, the activation functions \textit{ReLU} is adopted for all methods.

We trained all the aforementioned methods using PyTorch v1.3, where one Nvidia Titan RTX GPU with 24 GB memory is used. We used the L2 regularization with a decaying rate of $1\times10^{-3}$ and the Adam optimizer with a learning rate of $1\times10^{-3}$.
\subsection{Experimental Results}
\subsubsection{Learning-to-rank via Empirical Risk Minimization}
We note that the previous studies \cite{RevisitingApproxNDCG, StochasticTreatmentRF, LambdaLossFramework} just used a single fold (i.e., Fold1) for the experimental evaluation. To reduce the possible impact of overfitting on performance comparison, we use all the five folds and perform 5-fold cross validation. In particular, the dataset is randomly partitioned into five equal sized subsets. In each fold, three subsets are used as the training data, the remaining two subsets are used as the validation data and the testing data, respectively. We use the training data to learn the ranking model, use the validation data to select the hyper parameters based on nDCG@5, and use the testing data for evaluation. Finally, we report the ranking performance based on the averaged evaluation scores across five folds with $100$ epochs.

In order to show how the setting of activation function affects the performance of different learning-to-rank methods based on deep neural networks, 
we apply the same training framework for all the methods. Specifically, we used a simple $3$-layer feed-forward neural network, where the size of a hidden layer is set as $100$. Seven different activation functions are adopted, namely ReLU, LeakyReLU, RReLU, ELU, SELU, CELU and Sigmoid. Each method is evaluated with different activation functions, we report its best performance and the corresponding activation function in Table \ref{Table:ltraf}. We can observe that the optimal setting of activation function differs a lot. It reveals that it is necessary to carefully examine the setting of activation function when comparing different learning-to-rank methods or developing new methods. We note that it has been almost 14 years since the publication of LambdaRank. LambdaRank still achieves the best performance as shown in Table \ref{Table:ltraf}. This again reminds us that the open-source projects, such as PT-Ranking and TF-Ranking, are quite necessary for examining whether the reported improvements ``add up'' or not \cite{WorryingAnalysis}.

\begin{figure}[!htbp]
	\centering
	\includegraphics[width=2.8in, totalheight=1.8in]{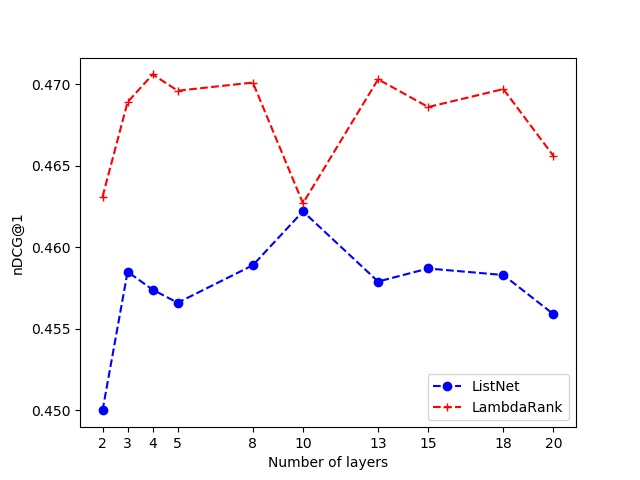}	
	\caption{The impact of number of layers on neural learning-to-rank.}
	\label{Fig:layer_effects}
\end{figure}

Furthermore, in Fig. \ref{Fig:layer_effects}, we plot the performance of ListNet and LambdaRank in terms of nDCG@1 with respect to the number of layers of the scoring function from $2$ to $20$. It is noticeable that the performance values of both ListNet and LambdaRank fluctuate when changing the number of hidden layers rather than a proportional improvement. One possible explanation is that: as the number of hidden layers increases, the ability of approximating more complex ranking functions (i.e., the model capacity) also increases. However, too many hidden layers may result in overfitting.

To summarize, the factors, such as different activation functions and  the number of layers, greatly affect the performance of a neural learning-to-rank method. Careful examinations of these factors are highly recommended in experimental comparisons of different learning-to-rank methods.

\begin{table}
	\caption{Performance of different learning-to-rank methods on MSLRWEB10K.}
	\label{Table:ltraf}
	\begin{centering}
		\resizebox{0.99\textwidth}{!}{
		\begin{tabular}{|l|c|c|c|c|c|c|c|}
			\hline 
			Method & Activation Function & nDCG@1 & nDCG@3 & nDCG@5 & nDCG@10 & nDCG@20 & nDCG@50\tabularnewline
			\hline 
			RankMSE & ReLU & 0.4469 & 0.4305 & 0.4328 & 0.4470 & 0.4693 & 0.5052\tabularnewline
			\hline 
			RankNet \cite{MSnDCGRankNet} & ELU & 0.4449 & 0.4346 & 0.4396 & 0.4557 & 0.4794 & 0.5142\tabularnewline
			\hline 
			LambdaRank \cite{LambdaRank} &  RReLU & \textbf{0.4670} & \textbf{0.4498} & \textbf{0.4528} & \textbf{0.4685} & \textbf{0.4910} & \textbf{0.5237}\tabularnewline
			\hline 
			ListNet \cite{ListNet} & ReLU & 0.4542 & 0.4324 & 0.4349 & 0.4500 & 0.4730 & 0.5075\tabularnewline
			\hline 
			ListMLE \cite{ListMLE} & ELU & 0.4523 & 0.4348 & 0.4395 & 0.4553 & 0.4767 & 0.5113\tabularnewline
			\hline 
			RankCosine \cite{RankCosine} & LeakyReLU & 0.4466 & 0.4300 & 0.4340 & 0.4487 & 0.4714 & 0.5073\tabularnewline
			\hline 
			ApproxNDCG \cite{QinApproximateNDCG} & Sigmoid & 0.4477 & 0.4263 & 0.4287 & 0.4428 & 0.4653 & 0.5000\tabularnewline
			\hline 
			WassRank \cite{TaoWSDM2019} & ELU & 0.4494 & 0.4306 & 0.4342 & 0.4494 & 0.4709 & 0.5059\tabularnewline
			\hline 
			ST-ListNet \cite{StochasticTreatmentRF} & ReLU & 0.4501 & 0.4346 & 0.4382 & 0.4532 & 0.4759 & 0.5111\tabularnewline
			\hline 
		\end{tabular}
	}
		\par\end{centering}
\end{table}

\subsubsection{Learning-to-rank via Adversarial Optimization}
Following the previous studies \cite{IRGAN, AdverSMIR}, we do not use the validation data when performing adversarial optimization. We use the training data to learn the ranking model, and the testing data for evaluation. Finally, we report the ranking performance based on the averaged evaluation scores across five folds with $100$ epochs.

In Table \ref{Table:perm}, we show the performance of adversarial learning-to-rank methods on MSLRWEB30K based on pointwise, pairwise, listwise generators and discriminators, respectively. As each method has two components, namely, generator and discriminator, we differentiate their performance with the suffixes (G) and (D), respectively. Moreover, the best result is indicated in bold.	From Table \ref{Table:perm}, we can observe that: (1) IRGAN-pair shows better performance than IRGAN-point, which echoes the experimental results in \cite{IRGAN}. (2) For pointwise, pairwise, listwise adversarial learning-to-rank methods, the discriminator ranking function achieves significantly better performance than the generator ranking function. (3) IRGAN-List is able to achieve better performance based on the discriminator. A possible reason is that the listwise discriminator considers the total order relationship among documents associated with the same query rather than the relative order between two documents treating documents independently. Furthermore, by increasing the size of ranking from $5$ to $10$, the listwise discriminator achieves slightly better performance.

\begin{table}[!htbp]
	\caption{Performance comparison on MSLRWEB30K.}
	\label{Table:perm}
	\begin{centering}
			\begin{tabular}{|l|c|c|c|c|}
				\hline 
				Method & nDCG@1 & nDCG@3 & nDCG@5 & nDCG@10\tabularnewline
				\hline 
				IRGAN-Point (D) & 0.2863 & 0.3019 & 0.3160 & 0.3447\tabularnewline
				\hline 
				IRGAN-Point (G) & 0.1658 & 0.1818 & 0.1963 & 0.2252\tabularnewline
				\hline 
				IRGAN-Pair (D) & 0.4254 & 0.4101 & 0.4150 & 0.4312\tabularnewline
				\hline 
				IRGAN-Pair (G) & 0.1929 & 0.2025 & 0.2123 & 0.2342\tabularnewline
				\hline 
				IRGAN-List-5 (D) & 0.4295 & 0.4133 & 0.4183 & 0.4347\tabularnewline
				\hline 
				IRGAN-List-5 (G) & 0.1549 & 0.1661 & 0.1785 & 0.2069\tabularnewline
				\hline 
				IRGAN-List-10 (D) & \textbf{0.4299} & \textbf{0.4141} & \textbf{0.4188} & \textbf{0.4350}\tabularnewline
				\hline 
				IRGAN-List-10 (G) & 0.1077 & 0.1137 & 0.1217 & 0.1408\tabularnewline
				\hline 
			\end{tabular}
		\par\end{centering}	
\end{table}

We note that MSLRWEB30K is a supervised dataset, where all the ground-truth labels of each training query are used during the optimization process. As reported by the prior studies \cite{IRGAN, AdverSMIR}, one potential advantage of adversarial learning-to-rank methods is the ability of allowing unlabelled documents within the training data. In order to understand well to what extent the ratio of unlabelled query-document pairs affects the performance of adversarial learning-to-rank, we randomly mask the ground-truth labels of each training query with a specific ratio. For instance, given the ratio of $0.2$, $20\%$ of ground-truth labels for each query will be masked as unlabelled. To reduce the possible impact of random masking and overfitting on performance comparison, we use all the five folds.

We show the performance of adversarial learning-to-rank methods on MSLRWEB30K with randomly masked labels in Table \ref{Table:maskperm}. This time the discriminators' performance is only reported, since generators always show poor performance as shown in Table \ref{Table:perm}. From Table \ref{Table:maskperm}, we can find that: (1) With the increase of unlabelled documents, both IRGAN-Point and IRGAN-Pair show decreased performance, which reveals that the performance is impacted with the increase of unlabelled documents. (2) On the contrary, IRGAN-List demonstrates robustness against the increase of unlabelled documents. This is attributable to the listwise sampling process when generating adversarial rankings.

\begin{table}[!htbp]
	\caption{Performance comparison in terms of nDCG@1 on MSLRWEB30K with randomly masked labels.}
	\label{Table:maskperm}
	\begin{centering}
			\begin{tabular}{|l|c|c|c|c|c|c|}
				\hline 
				Masking ratio & 0 & 0.1 & 0.2 & 0.3 & 0.4 & 0.5\tabularnewline
				\hline 
				IRGAN-Point (D) & 0.2863 & 0.2563 & 0.2725 & 0.2392 & 0.2532 & 0.2495\tabularnewline
				\hline 
				IRGAN-Pair (D) & 0.4254 & 0.4164 & 0.4154 & 0.4093 & 0.4078 & 0.3773\tabularnewline
				\hline 
				IRGAN-List-5 (D) & 0.4295 & 0.4324 & 0.4320 & 0.4345 & 0.4324 & 0.4317\tabularnewline
				\hline 
				IRGAN-List-10 (D) & \textbf{0.4299} & \textbf{0.4358} & \textbf{0.4371} & \textbf{0.4376} & \textbf{0.4368} & \textbf{0.4353}\tabularnewline
				\hline 
			\end{tabular}
		\par\end{centering}	
\end{table}
\section{Conclusion and Future Work}
In this work, we introduced PT-Ranking, an open-source package based on PyTorch. PT-Ranking is highly configurable for fine-tuning hyper-parameters and has easy-to-use APIs for developing new learning-to-rank models and optimization frameworks. PT-Ranking is highly complementary to the previous open-source projects for learning-to-rank. We envision that PT-Ranking will provide a convenient open-source platform for evaluating and developing learning-to-rank models based on deep neural networks, and thus facilitate researchers from different backgrounds.

For future work, first, we plan to add more learning-to-rank methods, such as \cite{LambdaLossFramework} and  \cite{AdverPRR}. Inspired by the recent studies \cite{nbdt, DeepNDF} on neural decision trees, it is interesting to include a number of learning-to-rank methods based on neural-backed decision trees. Second, we do note that the technique of neural architecture search (NAS) \cite{NASSurvey} can be applied for learning-to-rank. There is some hope that incorporating NAS will make PT-Ranking more versatile. Finally, we plan to add an interactive interface so that users can configure, evaluate and analyse learning-to-rank models in a visual manner.

\bibliographystyle{unsrt}



\begin{thebibliography}{10}
	
	\bibitem{SubsetRegression}
	David Cossock and Tong Zhang.
	\newblock Subset ranking using regression.
	\newblock In {\em Proceedings of the 19th Annual Conference on Learning
		Theory}, pages 605--619, 2006.
	
	\bibitem{RameshL2R}
	Ramesh Nallapati.
	\newblock Discriminative models for information retrieval.
	\newblock In {\em Proceedings of the 27th SIGIR}, pages 64--71, 2004.
	
	\bibitem{ChuGaussian}
	Wei Chu and Zoubin Ghahramani.
	\newblock Gaussian processes for ordinal regression.
	\newblock {\em Journal of Machine Learning Research}, 6:1019--1041, 2005.
	
	\bibitem{ChuSVOR}
	Wei Chu and {S.}~Sathiya Keerthi.
	\newblock New approaches to support vector ordinal regression.
	\newblock In {\em Proceedings of the 22nd ICML}, pages 145--152, 2005.
	
	\bibitem{FreundBoosting}
	Yoav Freund, Raj Iyer, Robert~E. Schapire, and Yoram Singer.
	\newblock An efficient boosting algorithm for combining preferences.
	\newblock {\em Journal of Machine Learning Research}, 4:933--969, 2003.
	
	\bibitem{ShenPerceptron}
	Libin Shen and Aravind~K. Joshi.
	\newblock Ranking and reranking with perceptron.
	\newblock {\em Machine Learning}, 60(1-3):73--96, 2005.
	
	\bibitem{RankSVMStruct}
	Thorsten Joachims.
	\newblock Training linear {SVMs} in linear time.
	\newblock In {\em Proceedings of the 12th KDD}, pages 217--226, 2006.
	
	\bibitem{RankCosine}
	Tao Qin, Xu{-}Dong Zhang, Ming{-}Feng Tsai, De{-}Sheng Wang, Tie{-}Yan Liu, and
	Hang Li.
	\newblock Query{-}level loss functions for information retrieval.
	\newblock {\em Information Processing and Management}, 44(2):838--855, 2008.
	
	\bibitem{OlivierMargin}
	Olivier Chapelle, Quoc Le, and Alex Smola.
	\newblock Large margin optimization of ranking measures.
	\newblock In {\em NIPS workshop on Machine Learning for Web Search}, 2007.
	
	\bibitem{AdaRank}
	Jun Xu and Hang Li.
	\newblock Adarank: a boosting algorithm for information retrieval.
	\newblock In {\em Proceedings of the 30th SIGIR}, pages 391--398, 2007.
	
	\bibitem{YueSVMAP}
	Yisong Yue, Thomas Finley, Filip Radlinski, and Thorsten Joachims.
	\newblock A support vector method for optimizing average precision.
	\newblock In {\em Proceedings of the 30th SIGIR}, pages 271--278, 2007.
	
	\bibitem{GuiverSoftRank}
	John Guiver and Edward Snelson.
	\newblock Learning to rank with softrank and gaussian processes.
	\newblock In {\em Proceedings of the 31st SIGIR}, pages 259--266, 2008.
	
	\bibitem{SoftRank}
	Michael Taylor, John Guiver, Stephen Robertson, and Tom Minka.
	\newblock Softrank{:} optimizing non-smooth rank metrics.
	\newblock In {\em Proceedings of the 1st WSDM}, pages 77--86, 2008.
	
	\bibitem{QinApproximateNDCG}
	Tao Qin, Tie{-}Yan Liu, and Hang Li.
	\newblock A general approximation framework for direct optimization of
	information retrieval measures.
	\newblock {\em Journal of Information Retrieval}, 13(4):375--397, 2010.
	
	\bibitem{LambdaMART}
	Qiang Wu, Christopher~J. Burges, Krysta~M. Svore, and Jianfeng Gao.
	\newblock Adapting boosting for information retrieval measures.
	\newblock {\em Journal of Information Retrieval}, 13(3):254--270, 2010.
	
	\bibitem{ListNet}
	Zhe Cao, Tao Qin, Tie{-}Yan Liu, Ming{-}Feng Tsai, and Hang Li.
	\newblock Learning to rank{:} from pairwise approach to listwise approach.
	\newblock In {\em Proceedings of the 24th ICML}, pages 129--136, 2007.
	
	\bibitem{ListMLE}
	Fen Xia, Tie{-}Yan Liu, Jue Wang, Wensheng Zhang, and Hang Li.
	\newblock Listwise approach to learning to rank{:} theory and algorithm.
	\newblock In {\em Proceedings of the 25th ICML}, pages 1192--1199, 2008.
	
	\bibitem{BoltzRank}
	Maksims~N{.} Volkovs and Richard~S{.} Zemel.
	\newblock Boltzrank{:} learning to maximize expected ranking gain.
	\newblock In {\em Proceedings of ICML}, pages 1089--1096, 2009.
	
	\bibitem{LambdaRank}
	Christopher~{J.C.} Burges, Robert Ragno, and Quoc~Viet Le.
	\newblock Learning to rank with nonsmooth cost functions.
	\newblock In {\em Proceedings of NeurIPS}, pages 193--200, 2006.
	
	\bibitem{IRGAN}
	Jun Wang, Lantao Yu, Weinan Zhang, Yu~Gong, Yinghui Xu, Benyou Wang, Peng
	Zhang, and Dell Zhang.
	\newblock Irgan{:} a minimax game for unifying generative and discriminative
	information retrieval models.
	\newblock In {\em Proceedings of the 40th SIGIR}, pages 515--524, 2017.
	
	\bibitem{AdverPRR}
	Xiangnan He, Zhankui He, Xiaoyu Du, and Tat{-}Seng Chua.
	\newblock Adversarial personalized ranking for recommendation.
	\newblock In {\em Proceedings of SIGIR}, pages 355--364, 2018.
	
	\bibitem{AdverSMIR}
	Dae~Hoon Park and Yi~Chang.
	\newblock Adversarial sampling and training for semi{-}supervised information
	retrieval.
	\newblock In {\em Proceedings of the Web Conference}, pages 1443--1453, 2019.
	
	\bibitem{AdverCMR}
	Bokun Wang, Yang Yang, Xing Xu, Alan Hanjalic, and Heng~Tao Shen.
	\newblock Adversarial cross{-}modal retrieval.
	\newblock In {\em Proceedings of International Conference on Multimedia}, pages
	154--162, 2017.
	
	\bibitem{AdverPreference}
	Zitai Wang, Qianqian Xu, Ke~Ma, Yangbangyan Jiang, Xiaochun Cao, and Qingming
	Huang.
	\newblock Adversarial preference learning with pairwise comparisons.
	\newblock In {\em Proceedings of International Conference on Multimedia}, pages
	656--664, 2019.
	
	\bibitem{AdverFGIMR}
	Kevin Lin, Fan Yang, Qiaosong Wang, and Robinson Piramuthu.
	\newblock Adversarial learning for fine{-}grained image search.
	\newblock In {\em Proceedings of ICME}, pages 490--495, 2018.
	
	\bibitem{WorryingAnalysis}
	Maurizio~Ferrari Dacrema, Paolo Cremonesi, and Dietmar Jannach.
	\newblock Are we really making much progress{?} a worrying analysis of recent
	neural recommendation approaches.
	\newblock In {\em Proceedings of the 13th ACM Conference on Recommender
		Systems}, pages 101--109, 2019.
	
	\bibitem{MetricRealityCheck}
	Kevin Musgrave, Serge Belongie, and Ser{-}Nam Lim.
	\newblock A metric learning reality check.
	\newblock {\em arXiv:2003.08505}, 2020.
	
	\bibitem{NeuralHype}
	Jimmy Lin.
	\newblock The neural hype and comparisons against weak baselines.
	\newblock In {\em SIGIR Forum}, pages 40--51, 2019.
	
	\bibitem{2015RIGOR}
	Jaime Arguello, Fernando Diaz, Jimmy Lin, and Andrew Trotman.
	\newblock Sigir 2015 workshop on reproducibility, inexplicability, and
	generalizability of results (rigor).
	\newblock In {\em Proceedings of the 38th SIGIR}, pages 1147--1148, 2015.
	
	\bibitem{PRML}
	Christopher~M. Bishop.
	\newblock {\em Pattern Recognition and Machine Learning}.
	\newblock Springer-Verlag, 2006.
	
	\bibitem{RankingSVM}
	Thorsten Joachims.
	\newblock Optimizing search engines using clickthrough data.
	\newblock In {\em Proceedings of the 8th KDD}, pages 133--142, 2002.
	
	\bibitem{MSnDCGRankNet}
	Chris Burges, Tal Shaked, Erin Renshaw, Ari Lazier, Matt Deeds, Nicole
	Hamilton, and Greg Hullender.
	\newblock Learning to rank using gradient descent.
	\newblock In {\em Proceedings of the 22nd ICML}, pages 89--96, 2005.
	
	\bibitem{TaoWSDM2019}
	Hai{-}Tao Yu, Adam Jatowt, Hideo Joho, Joemon Jose, Xiao Yang, and Long Chen.
	\newblock Wassrank{:} listwise document ranking using optimal transport theory.
	\newblock In {\em Proceedings of the 12th WSDM}, pages 24--32, 2019.
	
	\bibitem{StochasticTreatmentRF}
	Sebastian Bruch, Shuguang Han, Michael Bendersky, and Marc Najork.
	\newblock A stochastic treatment of learning to rank scoring functions.
	\newblock In {\em Proceedings of the 13th WSDM}, pages 61--69, 2020.
	
	\bibitem{LightGBM}
	Guolin Ke, Qi~Meng, Thomas Finley, Taifeng Wang, Wei Chen, Weidong Ma, Qiwei
	Ye, and Tie{-}Yan Liu.
	\newblock Lightgbm{:} {A} highly efficient gradient boosting decision tree.
	\newblock In {\em Proceedings of NeurIPS}, pages 3149--3157, 2017.
	
	\bibitem{XGBoost}
	Tianqi Chen and Carlos Guestrin.
	\newblock Xgboost{:} a scalable tree boosting system.
	\newblock In {\em Proceedings of the 22nd SIGKDD Conference}, pages 785--794,
	2016.
	
	\bibitem{RankEval}
	Claudio Lucchese, Cristina~Ioana Muntean, Franco~Maria Nardini, Raffaele
	Perego, and Salvatore Trani.
	\newblock Rankeval{:} an evaluation and analysis framework for
	{Learning-to-Rank} solutions.
	\newblock In {\em Proceedings of SIGIR}, pages 1281--1284, 2017.
	
	\bibitem{CatBoost}
	Liudmila Prokhorenkova, Gleb Gusev, Aleksandr Vorobev, Anna~Veronika Dorogush,
	and Andrey Gulin.
	\newblock Catboost{:} unbiased boosting with categorical features.
	\newblock In {\em Advances in Neural Information Processing Systems 32}, pages
	6638--6648, 2018.
	
	\bibitem{DSSM}
	Po-Sen Huang, Xiaodong He, Jianfeng Gao, Li~Deng, Alex Acero, and Larry Heck.
	\newblock Learning deep structured semantic models for web search using
	clickthrough data.
	\newblock In {\em Proceedings of CIKM}, pages 2333--2338, 2013.
	
	\bibitem{CDSSM}
	Yelong Shen, Xiaodong He, Jianfeng Gao, Li~Deng, and Gr{\'e}goire Mesnil.
	\newblock Learning semantic representations using convolutional neural networks
	for web search.
	\newblock In {\em Proceedings of the 23rd WWW}, pages 373--374, 2014.
	
	\bibitem{DRMM}
	Jiafeng Guo, Yixing Fan, Qingyao Ai, and W.~Bruce Croft.
	\newblock A deep relevance matching model for {Ad-hoc} retrieval.
	\newblock In {\em Proceedings of the 25th CIKM}, pages 55--64, 2016.
	
	\bibitem{HuCNNMatching}
	Baotian Hu, Zhengdong Lu, Hang Li, and Qingcai Chen.
	\newblock Convolutional neural network architectures for matching natural
	language sentences.
	\newblock In {\em Proceedings of 27th NIPS}, pages 2042--2050, 2014.
	
	\bibitem{PangAAAMatching}
	Liang Pang, Yanyan Lan, Jiafeng Guo, Jun Xu, Shengxian Wan, and Xueqi Cheng.
	\newblock Text matching as image recognition.
	\newblock In {\em Proceedings of AAAI Conference on Artificial Intelligence},
	pages 2793--2799, 2016.
	
	\bibitem{MatchSRNN}
	Shengxian Wan, Yanyan Lan, Jun Xu, Jiafeng Guo, Liang Pang, and Xueqi Cheng.
	\newblock Match{-}srnn{:} modeling the recursive matching structure with
	spatial rnn.
	\newblock In {\em Proceedings of IJCAI conference}, pages 2922--2928, 2016.
	
	\bibitem{TFRanking}
	Rama~Kumar Pasumarthi, Sebastian Bruch, Xuanhui Wang, Cheng Li, Michael
	Bendersky, Marc Najork, Jan Pfeifer, Nadav Golbandi, Rohan Anil, and Stephan
	Wolf.
	\newblock {TF-Ranking:} scalable {TensorFlow} library for {Learning-to-Rank}.
	\newblock In {\em Proceedings of KDD}, pages 2970--2978, 2019.
	
	\bibitem{MatchZoo}
	Jiafeng Guo, Yixing Fan, Xiang Ji, and Xueqi Cheng.
	\newblock Matchzoo{:} a learning, practicing, and developing system for neural
	text matching.
	\newblock In {\em Proceedings of SIGIR}, pages 1297--1300, 2019.
	
	\bibitem{GAN}
	Ian~J. Goodfellow, Jean Pouget{-}Abadie, Mehdi Mirza, Bing Xu, David
	Warde{-}Farley, Sherjil Ozair, Aaron Courville, and Yoshua Bengio.
	\newblock Generative adversarial nets.
	\newblock In {\em Proceedings of NeurIPS}, pages 2672--2680, 2014.
	
	\bibitem{LETORIR}
	Tao Qin, Tie-Yan Liu, Jun Xu, and Hang Li.
	\newblock {LETOR:} a benchmark collection for research on learning to rank for
	information retrieval.
	\newblock {\em Information Retrieval Journal}, 13(4):346--374, 2010.
	
	\bibitem{YahooL2RData}
	Olivier Chapelle and Yi~Chang.
	\newblock Yahoo{!} learning to rank challenge overview.
	\newblock In {\em Proceedings of the 2010 International Conference on YLRC},
	pages 1--24, 2010.
	
	\bibitem{FstnDCG}
	Kalervo J\"{a}rvelin and Jaana Kek\"{a}l\"{a}inen.
	\newblock Cumulated gain{-}based evaluation of {IR} techniques.
	\newblock {\em ACM Transactions on Information Systems}, 20(4):422--446, 2002.
	
	\bibitem{ERR}
	Olivier Chapelle, Donald Metlzer, Ya~Zhang, and Pierre Grinspan.
	\newblock Expected reciprocal rank for graded relevance.
	\newblock In {\em Proceedings of the 18th CIKM}, pages 621--630, 2009.
	
	\bibitem{RevisitingApproxNDCG}
	Sebastian Bruch, Masrour Zoghi, Michael Bendersky, and Marc Najork.
	\newblock Revisiting approximate metric optimization in the age of deep neural
	networks.
	\newblock In {\em Proceedings of the 42nd SIGIR}, pages 1241--1244, 2019.
	
	\bibitem{PlackettLuceModel-2}
	Robin~L. Plackett.
	\newblock The analysis of permutations.
	\newblock {\em Journal of the Royal Statistical Society{.} Series C},
	24(2):193--202, 1975.
	
	\bibitem{GumbelSoftmax}
	Eric Jang, Shixiang Gu, and Ben Poole.
	\newblock Categorical reparameterization with gumbel{-}softmax.
	\newblock In {\em International Conference on Learning Representations}, 2017.
	
	\bibitem{ConcreteDistribution}
	Chris~J. Maddison, Andriy Mnih, and Yee~Whye Teh.
	\newblock The concrete distribution{:} a continuous relaxation of discrete
	random variables.
	\newblock In {\em International Conference on Learning Representations}, 2017.
	
	\bibitem{LambdaLossFramework}
	Xuanhui Wang, Cheng Li, Nadav Golbandi, Michael Bendersky, and Marc Najork.
	\newblock The lambdaloss framework for ranking metric optimization.
	\newblock In {\em Proceedings of the 27th CIKM}, pages 1313--1322, 2018.
	
	\bibitem{nbdt}
	Alvin Wan, Lisa Dunlap, Daniel Ho, Jihan Yin, Scott Lee, Henry Jin, Suzanne
	Petryk, Sarah~Adel Bargal, and Joseph~E. Gonzalez.
	\newblock Nbdt: Neural-backed decision trees, 2020.
	
	\bibitem{DeepNDF}
	Peter Kontschieder, Madalina Fiterau, Antonio Criminisi, and Samuel~Rota Bulo.
	\newblock Deep neural decision forests.
	\newblock In {\em Proceedings of ICCV}, pages 1467--1475, 2015.
	
	\bibitem{NASSurvey}
	Thomas Elsken, Jan~Hendrik Metzen, and Frank Hutter.
	\newblock Neural architecture search{:} a survey.
	\newblock {\em Journal of Machine Learning Research}, 20(55):1--21, 2019.
	
\end{thebibliography}

\end{document}